\documentclass[a4paper, fleqn, usenatbib]{mnras}

\usepackage{mathptmx}
\usepackage{txfonts}

\usepackage[T1]{fontenc}
\usepackage{ae,aecompl}

\usepackage{graphicx}	
\usepackage{amssymb}	
\usepackage{color}
\usepackage{longtable}

\usepackage{color}

\definecolor{ao(english)}{rgb}{0.0, 0.5, 0.0}

\title[Detection of SGR~J1745$-$2900 up to 291$\,$GHz]{Detection of the magnetar SGR~J1745$-$2900 up to 291$\,$GHz with evidence of polarized millimetre emission}

\author[P.~Torne et al.]{P. Torne,$^{1}$\thanks{E-mail: ptorne@mpifr-bonn.mpg.de}
G.~Desvignes,$^{1}$
R.~P.~Eatough,$^{1}$
R.~Karuppusamy,$^{1}$ 
G.~Paubert,$^{3}$ \newauthor
M.~Kramer,$^{1,2}$
I.~Cognard,$^{4,5}$
D.~J.~Champion,$^{1}$
and
L.~G.~Spitler$^{1}$ \\
$^{1}$Max-Planck-Institut f\"{u}r Radioastronomie, Auf dem H\"{u}gel 69, D-53121, Bonn, Germany\\
$^{2}$Jodrell Bank Centre for Astrophysics, School of Physics and Astronomy, The University of Manchester, Manchester M13 9PL, UK\\
$^{3}$Instituto de Radioastronom\'{\i}a Milim\'{e}trica, Avda. Divina Pastora 7, N\'{u}cleo Central, 18012, Granada, Spain\\
$^{4}$Laboratoire de Physique et Chimie de l'Environnement et de l'Espace CNRS-Universit{\'e} d'Orl{\'e}ans, F-45071, Orl{\'e}ans, France\\
$^{5}$Station de radioastronomie de Nan{\c c}ay, Observatoire de Paris, CNRS/INSU, F-18330, Nan{\c c}ay, France
}

\date{Accepted 2016 October 21. Received 2016 October 17; in original form 2016 June 27}

\pubyear{2016}

\begin{document}
\label{firstpage}
\pagerange{\pageref{firstpage}--\pageref{lastpage}}
\maketitle


\begin{abstract}

In \citet{tor15}, we showed detections of SGR~J1745$-$2900 up to 225$\,$GHz (1.33$\,$mm); at that time the highest radio frequency detection of pulsar emission. In this work, we present the results of new observations of the same magnetar with detections up to 291$\,$GHz (1.03$\,$mm), together with evidence of linear polarization in its millimetre emission. 
SGR~J1745$-$2900 continues to show variability and is, on average, a factor $\sim$4 brighter in the millimetre band than in our observations of July 2014.
The new measured spectrum is slightly inverted, with $\left<\alpha\right> = +0.4\pm0.2$ (for $S_{\nu} \propto \nu^{\alpha})$. 
However, the spectrum does not seem to be well described by a single power law, which might be due to the intrinsic variability of the source, or perhaps a turn-up somewhere between 8.35 and 87$\,$GHz. These results may help us to improve our still incomplete model of pulsar emission and, in addition, they further support the search for and study of pulsars located at the Galactic Centre using millimetre wavelengths.

\end{abstract}




\section{Introduction}

Magnetar is the term used to refer to neutron stars whose high-energy luminosities can exceed their spin-down luminosity. These objects typically show large inferred magnetic fields ($B\,\rm{\gtrsim10^{13}\,G}$), and it is widely accepted that they require energy from magnetic field decay to power their emission, particularly at high energies \citep{dt92, td95, td96}.
The magnetars make up a small family within the pulsar population, with only 23 objects confirmed
\citep[see The Magnetar Catalog\footnote{\url{www.physics.mcgill.ca/~pulsar/magnetar/main.html}},][]{ok14}.
They are typically detected through their high-energy emission, but four of them have also shown radio pulsations \citep{cam06, cam07b, lev10, eat13}.

SGR~J1745$-$2900 is one of the radio emitting magnetars, and its location at the Galactic Centre, close to \mbox{Sgr A*} \citep{bow15}, makes it a particularly interesting object.
Studying the propagation effects of its emission can provide valuable information about the environment close to the supermassive black hole at the centre of the Galaxy and along the line-of-sight \citep[e.g.][]{eat13, sj13, bow14, spi14}.

The radio emission of magnetars is similar to that of the normal population of pulsars, but shows some remarkable differences.
For example, their flux density, spectral index, pulse profile shape, and polarization properties have been seen to vary on short and long time scales
\citep{cam06, cam07b, kra07, laz08, lev12, lynch15}. Such variability is inconsistent in most cases with propagation effects, and it is considered
intrinsic to the source.

Another peculiar characteristic of magnetar radio emission is the tendency to be spectrally flat, and SGR~J1745$-$2900 is no exception \citep{tor15}.
This is interesting because pulsars, being typically steep spectrum sources \citep[with mean spectral index $\rm{\left<\alpha\right>=-1.8}$,][]{mar2000}, are difficult to detect at radio frequencies above a few gigahertz. 
In fact, only seven normal pulsars have been detected above 30$\,$GHz to date \citep{wiele93, kra97b, morr97, loeh08}.
SGR~J1745$-$2900 held the record, prior to this work, with detections up to 225$\,$GHz \citep{tor15}, followed by XTE~J1810$-$197 up to 144$\,$GHz \citep{cam07c}.

The study of the characteristics of pulsar radio emission at high frequencies can help to elucidate how the emission from neutron stars is produced;
a problem that remains unresolved since the discovery of pulsars almost 50 years ago \citep[see e.g.][]{hank09, melrose2016}.
For instance, some models predict a possible turn-up in the spectrum at sufficiently high frequencies, due to incoherent emission becoming dominant \citep{mich78, mich82}.
Observational works have reported an excess of flux density for some of the pulsars studied at millimetre wavelengths \citep{wiele93, kra96, kra97b}, giving credibility to those models.
However, the sample of pulsar observations at high radio frequencies is small. More observations are needed to study better the behaviour of the high frequency emission, and to check for turn-ups or other unpredicted effects.

Because of its high luminosity and flat spectrum \citep{tor15}, SGR~J1745$-$2900 is a superb source to be observed at very high radio frequencies, especially at the short millimetre regime where there is almost no information about pulsar radiation. This work presents the results from a multifrequency campaign for SGR~J1745$-$2900 carried out at frequencies between 2.54 and 291$\,$GHz (wavelengths between 11.8$\,$cm and 1.03$\,$mm) aiming to obtain further information about its emission properties and providing additional constraints for pulsar emission models.

\section{Observations and Data Analysis}\label{sec:obs_analysis}

The millimetre observations were made with the 30-m radio telescope of the Institut de Radioastronomie Millim\'{e}trique (IRAM) during 2015 March 4-9. The receiver used was the Eight MIxer Receiver \citep[EMIR,][]{car12}. EMIR delivers four separated, tunable frequency bands between $\sim$73 and 350$\,$GHz (4 and $\rm{0.8\,mm}$) in dual linear polarization\footnote{For more information on the EMIR frequency combinations, see \url{http://www.iram.es/IRAMES/mainWiki/EmirforAstronomers}}. During the six days of observations different set-ups were used, mainly depending on weather, covering between 87 and 291$\,$GHz. The Broad-Band-Continuum (BBC) backend recorded the four bands tuned in EMIR with $\rm{\sim6\,GHz}$ of bandwidth each ($\rm{\sim24\,GHz}$ in total), no frequency resolution (i.e. total power mode), with a sampling time of $\rm{1\,ms}$. After several upgrades, the intermediate frequency (IF) range of all EMIR mixers is 4 to 12 GHz which is directly fed to the BBC power detectors. 
However, there is a significant slope in the passband, which favours the lower 
frequencies, and we have therefore taken 6$\,$GHz (instead of 8$\,$GHz) as the properly weighted effective IF bandwidth. 
The sky frequency values corresponding to the centre of the 4$-$12$\,$GHz IF range of the used EMIR mixers are 87, 101, 138, 154, 209, 225, 275, and 291$\,$GHz.

In addition to the observations with the \mbox{IRAM~30-m}, we observed simultaneously at certain epochs with the \mbox{Effelsberg~100-m} radio telescope and with the \mbox{Nan\c{c}ay 94-m} equivalent radio telescope. At Effelsberg, two different observing frequencies were used, centred at 4.85 and $\rm{8.35\,GHz}$.
At Nan\c{c}ay, the central frequency was 2.54$\,$GHz. The set-up, data reduction and calibration of the Effelsberg and Nan\c{c}ay data were identical to those already described in \citet{tor15}. Table \ref{table:obs_summary} summarizes the observations.

At \mbox{IRAM}, each observing session consisted typically of 45-min scans on SGR~J1745$-$2900 with  interspersed ``hot-cold-sky'' calibration measurements. Additionally, a few scans on planets were used to verify the absolute flux density calibration, obtained following the methodology in \citet{ckra97}, and applying elevation and frequency-dependent gain corrections \citep{pen12}. 

Typical ``switching techniques'' used in millimetre observations to subtract the atmospheric contribution are not adequate for pulsar observations. Thus, the time series at millimetre wavelengths required careful processing to enhance the detections of the magnetar. This is because of a significant amount of red noise present in the data, mostly due to variations in the atmospheric water vapour content during the observations. Such effect can be particularly bad for observations of SGR~J1745$-$2900 due to its long period and the low elevation at which the \mbox{IRAM~30-m} sees the Galactic Centre (elevation$\,<\,$25$\,$deg), which translates into a considerable 
airmass\footnote{Airmass refers to the amount of Earth's atmosphere that a celestial signal passes through along the line of sight.}. The data showed also periodic interference, the most prominent being at 1 and 50$\,$Hz and some of their harmonics, most likely related to the cryogenerator and the mains power.

The cleaning process was as follows. First, the time series was Fourier transformed and prominent interference was removed by zapping a few spectral bins around each peak at 1, 2, 50, 55, 60, 100 and 200$\,$Hz. After an inverse Fourier transform, the resulting time series was filtered by a running mean with a window length of 10$\,$s. To prevent the filtering from degrading the magnetar pulses, we used an ephemeris from a timing model to predict the times of arrival of the pulses at the observatory, and protected a window of 0.81$\,$s (approximately the pulse width) around each time-of-arrival. This was achieved by extrapolating 3.75$\,$s of the running mean vector from each side of the protection window by using a third-degree polynomial. For the two highest observing frequencies, 275 and 291$\,$GHz, we extrapolated a shorter block of running mean vector of only 0.05$\,$s to each side, using a linear interpolation, which gave better results. Once we had a modified time series with the content of the 
protected windows substituted by the extrapolated data, we apply a running mean on this new time series that is then subtracted from the original one. This method is effective at cleaning the pulse, while avoiding the artefacts caused by the running mean. Next, we removed some negative spikes that occasionally appear in the time series by substituting all negative values of the time series larger than $-6\sigma$ with the median of the time series. A final step was to apply a second running mean filter with a smaller window of 0.4$\,$s, again protecting the pulse window, to remove short-term variations. 

The result of the cleaning process is a high signal-to-noise folded profile with nearly flat off-pulse baseline. The mean flux density (i.e. the integrated profile intensity averaged over the full period) is calculated by summing the area under the pulse and dividing by the number of bins in the profile. Flux density errors are estimated from the off-pulse noise and also include the uncertainties in the absolute calibration factors, estimated to 10, 20 and 30 per cent for the 3, 2 and 1$\,$mm bands, respectively.

\section{Results and Discussion}

\subsection{Detections, flux density and spectrum}

SGR~J1745$-$2900 was detected at all frequencies from 2.54 up to 291$\,$GHz. The weather conditions at \mbox{IRAM~30-m} were excellent during five days of observations (zenith opacity~at~225$\,$GHz~<~0.25), contributing to the successful detections at the highest frequencies. In addition, the magnetar was particularly bright in the millimetre band during this observing campaign, with a varying flux density of mean value (averaged over all the millimetre observations) $\rm{\left<\emph{S}_{mm} \right> = 5.5\,\pm\,0.4\,mJy}$.
This is a factor $\sim$4 brighter than in July 2014 \citep{tor15}. In contrast, at the lowest frequencies, 2.54 to 8.35 GHz, the magnetar was a factor $\sim$6 dimmer than in July 2014.

\begin{figure}
    \begin{center}
   \includegraphics[width=0.95\linewidth]{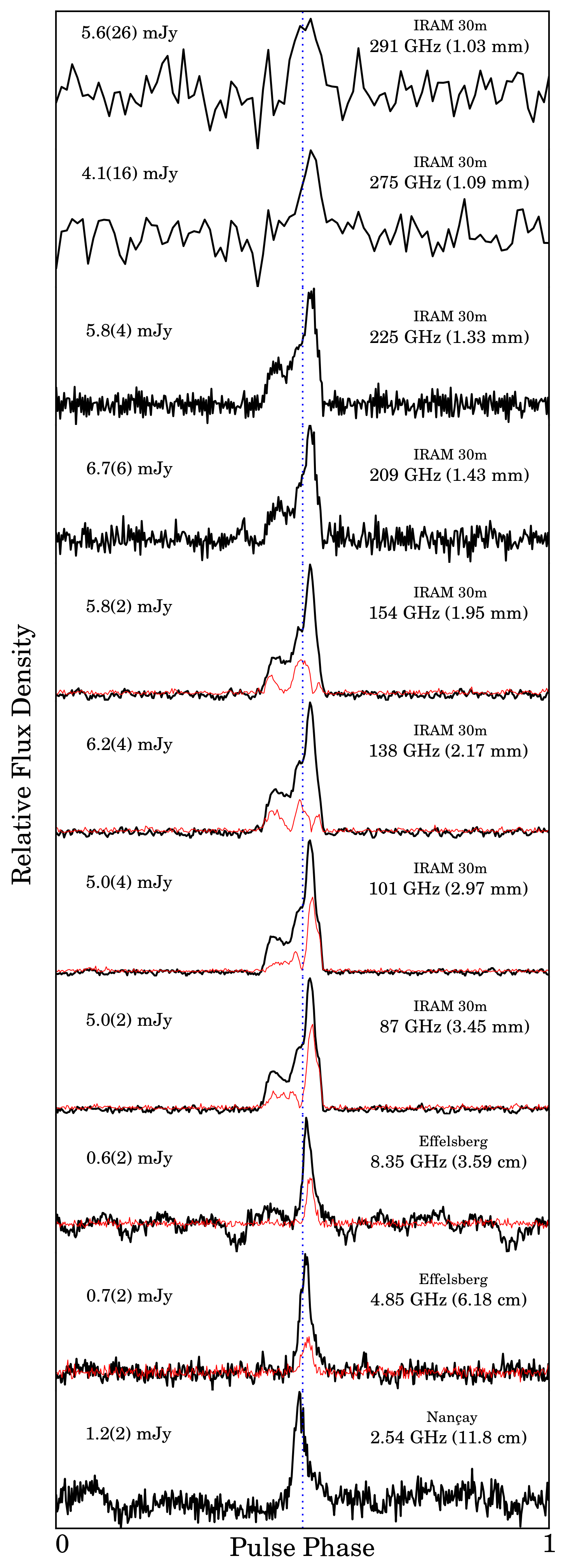}
    \caption{Average profiles of SGR~J1745$-$2900 from 2.54 up to 291$\,$GHz. 
    The black thick line represents the total intensity profile, and the red thinner line shows the linear polarization, which is a lower limit between 87 and 154$\,$GHz (see text).
    The vertical dotted line marks the predicted rotational phase from the ephemeris.
    }
    \label{fig:HFprofiles}
    \end{center}
\end{figure}

\begin{figure}
    \begin{center}
   \includegraphics[width=\linewidth]{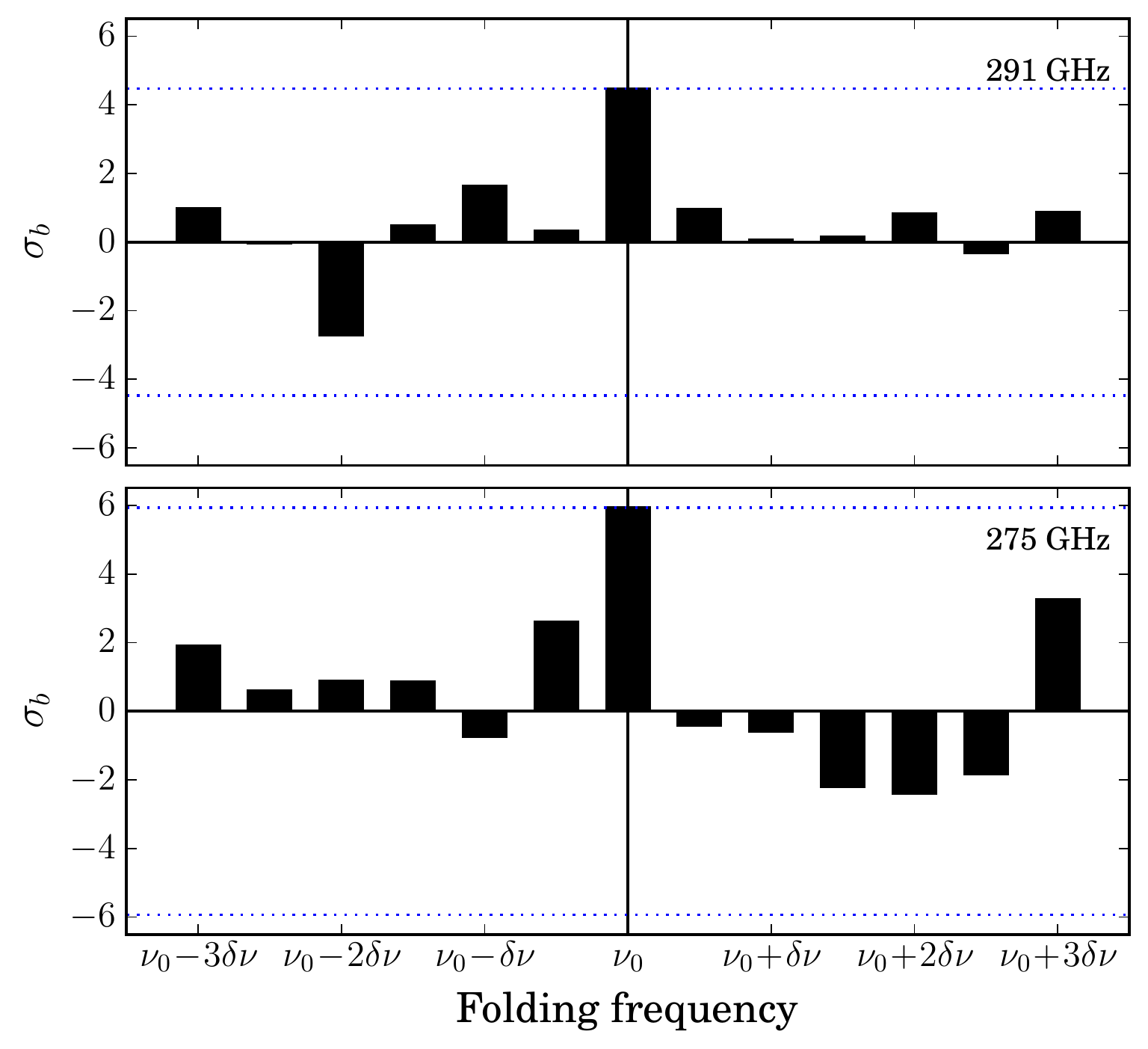}
    \caption{Periodogram of the significance of the profiles at 275 and 291$\,$GHz when applying the cleaning method without protecting the on-pulse window from the running mean. The significance figure is calculated by adding the intensity over the on-pulse window and dividing by the standard deviation of the off-pulse region: $\sigma_\mathrm{b} = \sum I[\mathrm{on}] / \sqrt{ \sum (I[\mathrm{off}]-\mu)^2 / \mathrm{N} }$, where $\mu$ and $\mathrm{N}$ are, respectively, the mean and the number of bins in the off-pulse region. The folding frequency is shifted in steps of half a Fourier bin of the longest individual integration length (45$\,$min), i.e. $\delta \nu = 0.370\,$mHz. Each step therefore corresponds to a variation of the folding frequency of $\approx$0.07 per cent. $\sigma_b$ reaches values of 5.9 and 4.5, for 275 and 291$\,$GHz, respectively; values that increase to $\sigma_b=$18.5 and 10.6, respectively, when the pulse is protected from the running mean as described in 
    Section~\ref{sec:obs_analysis}.
    }
    \label{fig:periodogram}
    \end{center}
\end{figure}

Figure \ref{fig:HFprofiles} shows the averaged pulse profiles of SGR~J1745$-$2900 at all frequencies from 2.54 to 291$\,$GHz. The detections at 209 and 225$\,$GHz are now clear with peak signal-to-noise ratios of about 12, confirming the tentative detections presented in \citet{tor15}. At the highest frequencies, 275 and 291$\,$GHz, the detections are weaker, but the alignment of the peaks with the pulse arrival phase predicted by the timing ephemeris and the simultaneous detections at other frequencies strengthen their significance. To verify that the pulses at 275 and 291~GHz were not an artefact of the on-pulse window protection in our cleaning method we carried out several tests. First, we applied the same cleaning shifting the protection window by 0.25 in rotational phase, and confirmed that no peak was artificially produced. Secondly, we applied the cleaning method without the on-pulse window protection, obtaining also peaks, although less significant. This is expected since the running mean tends to 
dip the pulse if not protected. We produced a periodogram of the significance of the profiles at 275 and 291$\,$GHz without protecting the on-pulse window, folding the data at the spin frequency of the magnetar, and at nearby frequencies. The periodogram shows that the significant pulses are present only when folding at exactly the correct spin frequency (see Fig.~\ref{fig:periodogram}).

Table \ref{table:fluxes} presents the measured mean flux densities and spectral index per day, together with the total averaged values. The high system temperatures and red noise (dominated by atmospheric effects) at 275 and 291$\,$GHz made the detections at these frequencies challenging on individual days. 
Once the observations were combined, we obtained more significant detections (see Fig. \ref{fig:HFprofiles}) and measurements of the mean flux densities. 

We remark that the flux density values presented in Table \ref{table:fluxes} for each frequency are averages per day. In some cases, the intensity of SGR~J1745$-$2900 varied between different single observations by up to a factor of 2, in less than a few hours. Furthermore, within single observations, the flux density is sometimes seen to vary by a factor of a few in what seems to be a bursty behaviour. We investigated if the scintillation effect in the ISM could be responsible for this variability. Following \citet{cl91}, we calculate that the scintillation at the lowest frequencies, 2.54, 4.85 and 8.35$\,$GHz, is negligible (and so, the variability of the magnetar at these frequencies must be intrinsic). At the millimetre wavelengths (87$\,$GHz and above) the scintillation cannot be fully neglected, but at most it could account for an intensity modulation of a few tens of per cent. This could explain small intensity variations between consecutive days (see Table \ref{table:fluxes}), as the refractive 
interstellar scintillation can have time scales down to a few days, but it cannot account for the variations of factors of a few that we also observe. Therefore, at the millimetre wavelengths the emission of SGR~J1745$-$2900 must also have a large fraction of intrinsic variability. Furthermore, the variations of SGR~J1745$-$2900 are not only in radio flux density and spectral index, but also in profile shape and polarization characteristics, a behaviour similar to what has been reported for this and the other known radio magnetars \citep[e.g.][]{kra07, cam07a, cam08, lev12, lynch15, pen15}.

\begin{figure}
    \begin{center}
    \includegraphics[width=\linewidth]{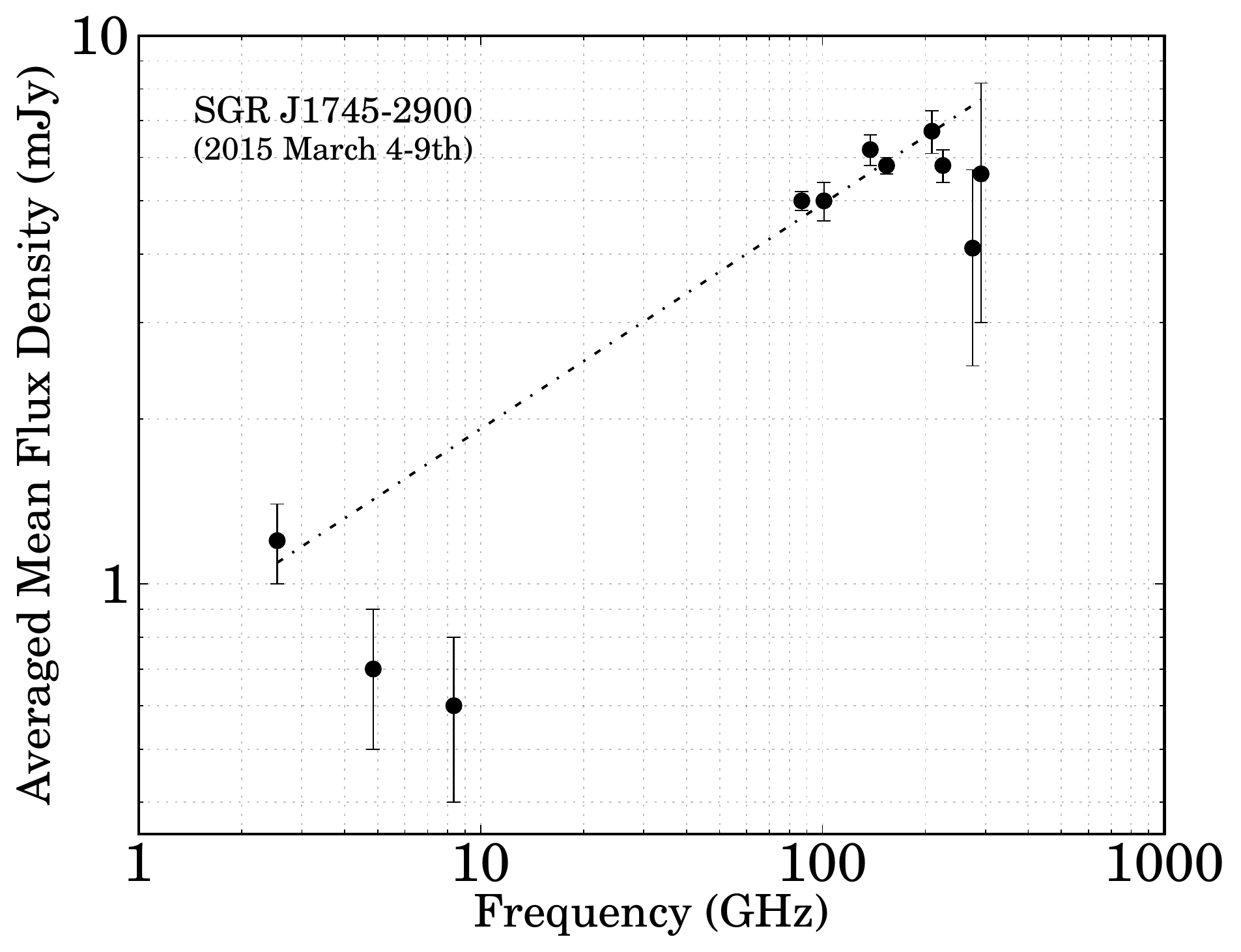}
    \caption{Average spectrum of SGR~J1745$-$2900 from the observations. The dashed-dotted line shows the spectral index fit using a single power law.
    The mean spectral index obtained is $\rm{\left<\alpha\right>\,=\,+0.4\,\pm\,0.2}$. Error bars are 2$\sigma$.
    }
    \label{fig:MAR15spectrum}
    \end{center}
\end{figure}

Figure \ref{fig:MAR15spectrum} shows the observed averaged spectrum of SGR~J1745$-$2900. Interestingly, the magnetar was clearly weaker at the lower frequencies, 2.54 to 8.35$\,$GHz. A single power law fit yields a slightly inverted spectral index of $\rm{\left<\alpha\right>}=+0.4\pm0.2$. This is still consistent with a flat spectrum typical from radio loud magnetars, but it is somewhat different to the spectrum observed for the same source in July 2014 \citep[$\rm{\left<\alpha\right>}=-0.4\pm0.2$;][]{tor15}. 

It is noticeable that the single power law does not fit well all the data points (in particular 4.85 and 8.35$\,$GHz). We can think of several possible explanations.
The first is that the intrinsic variability of the source behaves differently at different frequencies and deviates the spectrum from the single power law. 
A second possibility is that the measurements at low frequencies suffer from some systematic error, perhaps due to interference or red noise. The detections at Effelsberg were weaker than in the previous observing campaign in July 2014 \citep{tor15} and the profile baseline showed some non-Gaussian noise (see e.g. the 8.35$\,$GHz averaged profile in Fig. \ref{fig:HFprofiles}), which could lead to an under or overestimation of the flux density if the pulse lies on a dip or bump of the baseline, respectively. However, such effects would not account for deviations in the measurements larger than a few tens of per cent and are reflected in the errors at those frequencies. Thus, we consider this explanation less likely. 
Finally, a last possible explanation could be that there is a turn-up in the spectrum somewhere between 8.35 and 87$\,$GHz. A turn-up might occur as a result of the decrease in the efficiency of the coherent radiation mechanism, together with an incoherent component of emission that could take over. This effect is predicted by some pulsar emission models \citep{mich78, mich82}, and hints of turn-ups have in fact been observed in some pulsars at around $\sim$30$\,$GHz \citep{wiele93, kra96, kra97b}.
Moreover, an incompatibility with a single power law has also been reported for the spectrum of another radio magnetar, \mbox{1E~1547.0$-$5408} \citep{cam08}, suggesting the possibility of more complex spectra in radio magnetars than the typical single or broken power law of normal pulsars. The location of the possible turn-up in pulsar emission is not clear \citep[it could be somewhere between radio and infrared, see][]{mich82}, and SGR~J1745$-$2900 is at the moment the only pulsar detectable from a few up to a few hundreds of gigahertz, which could be key to detect that possible turn-up in its spectrum. Additional simultaneous multifrequency observations covering the region around $\sim$30$-$40$\,$GHz would be helpful to solve the turn-up question.

\begin{table*}
  \centering
  \caption{Summary of the observations. For each day and frequency ($\rm{\nu}$), the total integration time ($\rm{Tobs_{\nu}}$) on SGR~J1745-2900 is given in minutes. The symbol ``$-$'' means that no observation was done at that frequency on that particular day. The observations at 2.54$\,$GHz were taken with Nan\c cay, 4.85 and 8.35$\,$GHz with Effelsberg, and 87 to 291$\,$GHz
  with \mbox{IRAM~30-m}.
  }
  \label{table:obs_summary}
  \begin{tabular}{@{}ccccccccccccc@{}}
  \hline
   Date		& $\rm{Tobs_{2.54}}$ & $\rm{Tobs_{4.85}}$ & $\rm{Tobs_{8.35}}$ & $\rm{Tobs_{87}}$ & $\rm{Tobs_{101}}$ & $\rm{Tobs_{138}}$ & $\rm{Tobs_{154}}$ & $\rm{Tobs_{209}}$ & $\rm{Tobs_{225}}$ & $\rm{Tobs_{275}}$ & $\rm{Tobs_{291}}$ & MJD\\
		&  (min)      & (min)     & (min)      & (min)            & (min)    & (min)     & (min)     & (min)     & (min)     & (min)     & (min)     & (days)\\
 \hline
 2015 Mar 04 & $55$   & $72$ & $72$ & $180$ & $180$ & $90$  & $90$   & $90$ & $90$ & $-$  & $-$  & 57085  \\
 2015 Mar 05 & $-$    & $-$  & $-$  & $90$  & $90$  & $90$  & $90$   & $90$ & $90$ & $90$ & $90$ & 57086  \\
 2015 Mar 06 & $-$    & $60$ & $60$ & $90$  & $90$  & $90$  & $90$   & $90$ & $90$ & $90$ & $90$ & 57087  \\
 2015 Mar 07 & $72$   & $-$  & $-$  & $78$  & $78$  & $45$  & $45$   & $78$ & $78$ & $45$ & $45$ & 57088  \\
 2015 Mar 08 & $-$    & $66$ & $72$ & $90$  & $90$  & $185$ & $185$  & $-$ & $-$ & $95$ & $95$ & 57089  \\
 2015 Mar 09 & $72$   & $-$  & $-$  & $45$  & $45$  & $45$  & $45$   & $-$  & $-$  & $-$  & $-$  & 57090  \\
 \hline
\end{tabular}
\end{table*}

\begin{table*}
  \centering
  \caption{Measured flux densities and spectral indices of SGR~J1745$-$2900. Two-sigma errors in the last digits are shown in parentheses. The symbol ``$-$'' means that no observation was done at that frequency on that particular day. ``$\rm{ND}$'' indicates observations with no detection.
  }
  \label{table:fluxes}
  \begin{tabular}{@{}ccccccccccccc@{}}
  \hline
   Date		& $S_{2.54}$ & $S_{4.85}$ & $S_{8.35}$ & $S_{87}$ & $S_{101}$ & $S_{138}$ & $S_{154}$ & $S_{209}$ & $S_{225}$ & $S_{275}$ & $S_{291}$ & $\alpha$ \\
   (2015)	& (mJy)     & (mJy)      & (mJy)       & (mJy)    & (mJy)     & (mJy)     & (mJy)     & (mJy)     & (mJy)     & (mJy)     & (mJy)     &          \\
 \hline
 2015 Mar 04   & $1.0(2)$   & $0.9(2)$   & $0.8(2)$    & $6.2(2)$ & $6.4(2)$ & $7.7(2)$  & $6.3(4)$   & $5.9(14)$ & $3.7(8)$ & $-$          & $-$          &  $+0.5(4)$ \\
 2015 Mar 05   & $-$        & $-$        & $-$         & $5.4(2)$ & $5.0(2)$ & $6.3(2)$  & $5.3(2)$   & $6.9(8)$  & $7.0(6)$ & $4.6(28)$    & $4.5(38)$    &  $+0.2(6)$ \\
 2015 Mar 06   & $-$        & $0.7(2)$   & $0.6(2)$    & $3.7(2)$ & $3.7(2)$ & $7.5(2)$  & $6.5(2)$   & $5.5(10)$ & $5.1(8)$ & $5.9(24)$    & $6.7(36)$    &  $+0.8(6)$ \\
 2015 Mar 07   & $1.4(6)$   & $-$        & $-$         & $6.4(2)$ & $5.7(2)$ & $6.4(2)$  & $5.8(2)$   & $8.5(12)$ & $7.2(6)$ & $\rm{ND}$ & $\rm{ND}$ &  $+0.1(6)$ \\
 2015 Mar 08   & $-$        & $0.4(2)$   & $0.4(2)$    & $5.1(2)$ & $5.1(2)$ & $5.3(2)$  & $5.9(2)$   & $-$       & $-$      & $1.6(30)$    & $\rm{ND}$ &  $+0.4(6)$ \\
 2015 Mar 09   & $1.2(2)$   & $-$        & $-$         & $3.4(2)$ & $4.3(4)$ & $3.8(6)$  & $4.7(6)$   & $-$       & $-$      & $-$          & $-$          &  $+0.3(2)$ \\
 Total Average & $1.2(2)$   & $0.7(2)$   & $0.6(2)$    & $5.0(2)$ & $5.0(4)$ & $6.2(4)$  & $5.8(2)$   & $6.7(6)$  & $5.8(4)$ & $4.1(16) $   & $5.6(26)$    &  $+0.4(2)$ \\
 \hline
\end{tabular}
\end{table*}

\subsection{Linear polarization}

\begin{figure}
    \begin{center}
    \includegraphics[width=\linewidth]{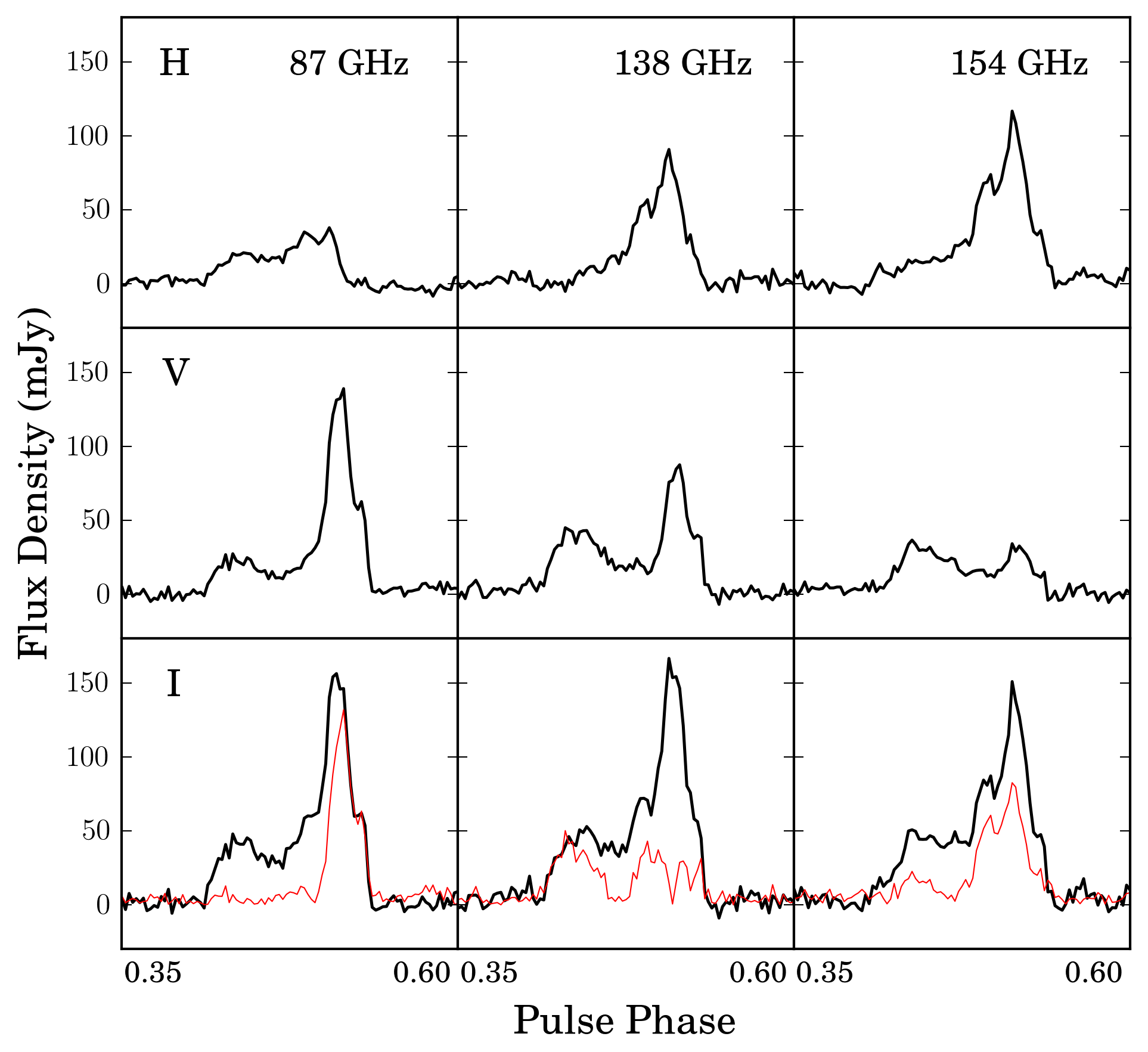}
    \caption{Selected examples of the pulse profile morphology differences seen in the two linear feeds of the \mbox{IRAM~30-m} at three different frequencies. The top panels show the profile detected on the horizontal feed (H), the middle panels on the vertical feed (V), and the bottom panels show the total intensity $\rm{I=H+V}$ (black thick line), with the lower limit of the degree of linear polarization (red thin line, see text). The differences in the profile shape between the two feeds are recurrent, and the lower limit on the degree of linear polarization is greater than zero in many bins across the profile (see also Fig. \ref{fig:HFprofiles}), reaching values of up to 100 per cent linearly polarized emission for some profile bins.     
    }
    \label{fig:POLcomp}
    \end{center}
\end{figure}

Apart from the total intensity detections, we observe evidence of linear polarization in the emission from SGR~J1745$-$2900, including at the millimetre wavelengths. This is obvious from the comparison of the pulse profile morphology from the horizontal (H) and vertical (V) linear feeds of the \mbox{IRAM~30-m}, which clearly show a different profile shape in most of the observations. 
Figure \ref{fig:POLcomp} shows a comparison of the H and V pulse profiles at 87, 101 and 154$\,$GHz as an example. 

This pulse profile morphology inequality when recording data with orthogonal linear feeds is indicative of radiation that must have a certain degree of linear polarization. Unfortunately, the BBC backend does not provide all Stokes parameters, making it not possible to quantify the degree of both linear and circular polarization in the millimetre emission directly. However, we can calculate Stokes Q from the power of the two linear feeds and set a lower limit in the degree of linear polarization: $L = \sqrt{Q^2+U^2} \geq |Q|$. The \emph{lower limit} of $L$ reaches values as high as 100 per cent for certain profile bins, and it is greater than zero for most of the pulse (see Fig. \ref{fig:POLcomp} and \ref{fig:HFprofiles}). Moreover, minimum values of $L$ greater than zero are noticeable in certain profile bins up to 225$\,$GHz, and tentatively at 275$\,$GHz. At 291$\,$GHz the detections are too weak to be conclusive. However, the noise levels are higher above 154$\,$GHz, and we also note apparent correlated 
noise between the two polarizations for frequencies between 209 and 291$\,$GHz. The origin of this noise is not fully understood and, for this reason, we do not show here the lower limits on $L$ for frequencies above 154$\,$GHz. Ongoing work for a more detailed analysis of the polarized emission from SGR~J1745$-$2900 will be presented in a future publication (Wucknitz et al. in prep.). 

This is the first time linearly polarized emission up to 154$\,$GHz (1.95$\,$mm) has been seen in a pulsar. It is closely followed by XTE~J1810$-$197, with inferred linear polarization up to 144$\,$GHz \citep{cam07c}. For normal pulsars, the highest radio frequency at which polarization has been detected is 32$\,$GHz \citep{xil96}. Polarized emission at such high level and frequencies as seen for SGR~J1745$-$2900 is unusual, as radio pulsars have been reported to depolarize at high radio frequencies \citep{morr1981, xil96}. In contrast, radio magnetars can stay highly linearly polarized up to very high frequencies \citep{kra07, cam08}. Consequently, the polarized emission at millimetre wavelengths of SGR~J1745$-$2900 is not totally unexpected. In fact, \citet{krav16} measured an averaged linear polarization for this magnetar of about 65 per cent at $\sim$40$\,$GHz. 

The polarization in pulsar emission is generally linked with a coherent radiation mechanism. The fact that high degrees of linear polarization are measured at the
millimetre wavelengths for SGR~J1745$-$2900 therefore may conflict and weaken the idea of an incoherent component of emission responsible for a turn-up in its spectrum.

\section{Summary}


We show in this paper that the radio emission from highly magnetized neutron stars can reach extremely high frequencies with detections that reach 291$\,$GHz (1.03$\,$mm). These new detections break the previous record recently set in \citet{tor15} as the highest radio frequency detection of pulsar emission, and give us more hints about the radiation mechanism of these objects. SGR~J1745$-$2900 continues to show significant variability in its emission characteristics, and its averaged mean flux density in the millimetre band is a factor $\sim$4 higher than in July 2014, while between 2.54 and 8.35$\,$GHz is on average a factor $\sim$6 dimmer.
Furthermore, we show evidence for a significant degree of linear polarization in the millimetre emission from SGR~J1745$-$2900, reaching factors up to a 100 per cent for certain profile bins, and being the polarized emission at the highest radio frequencies ever detected from a pulsar. The measured spectrum is slightly inverted, with a spectral index of $\left<\alpha\right> = +0.4\pm0.2$ when a single power law is fit. The spectrum has an uncommon shape, with decaying flux density between 2.54 and 8.35$\,$GHz and a much stronger emission at the millimetre band, which may be due to intrinsic intensity variability or indicative of the existence of a turn-up in the emission somewhere between 8.35 and 87$\,$GHz. These new results are relevant to the development of better pulsar emission models, in particular those trying to explain the radio emission from magnetars; and are also further proof that we can detect and study pulsars located at the Galactic Centre using millimetre wavelengths.

\section*{Acknowledgements}

We thank the anonymous referee for a careful review and constructive 
comments that helped improving the manuscript.
We also thank Olaf Wucknitz and Dominic Schnitzeler 
for discussions on the polarization analysis, Jim Cordes
for discussions on scintillation effects,
and the staff at the \mbox{IRAM~30-m} for their great support.
Based on observations carried out with the \mbox{IRAM~30-m},
the \mbox{Effelsberg~100-m}, and the Nan{\c c}ay radio telescopes. 
The Nan{\c c}ay radio observatory is operated by the Paris Observatory,
associated to the French CNRS.
The \mbox{Effelsberg~100-m} is operated by the MPIfR (Max-Planck-Institut f\"ur Radioastronomie).
IRAM is supported by INSU/CNRS (France), MPG (Germany) and IGN (Spain).
P.T. is supported for this research through a stipend from the
International Max Planck Research School (IMPRS).
Financial support by the European Research Council for the
ERC SynergyGrant \emph{BlackHoleCam} (ERC-2013-SyG, Grant Agreement
no. 610058) is gratefully acknowledged.
L.G.S. gratefully acknowledges financial support from the ERC Starting Grant BEACON under contract no. 279702.
  



\bibliographystyle{mnras}
\bibliography{Torne_etal_Pico_J1745-2900_291GHzPol} 

\begin{thebibliography}{}
\makeatletter
\relax
\def\mn@urlcharsother{\let\do\@makeother \do\$\do\&\do\#\do\^\do\_\do\%\do\~}
\def\mn@doi{\begingroup\mn@urlcharsother \@ifnextchar [ {\mn@doi@}
  {\mn@doi@[]}}
\def\mn@doi@[#1]#2{\def\@tempa{#1}\ifx\@tempa\@empty \href
  {http://dx.doi.org/#2} {doi:#2}\else \href {http://dx.doi.org/#2} {#1}\fi
  \endgroup}
\def\mn@eprint#1#2{\mn@eprint@#1:#2::\@nil}
\def\mn@eprint@arXiv#1{\href {http://arxiv.org/abs/#1} {{\tt arXiv:#1}}}
\def\mn@eprint@dblp#1{\href {http://dblp.uni-trier.de/rec/bibtex/#1.xml}
  {dblp:#1}}
\def\mn@eprint@#1:#2:#3:#4\@nil{\def\@tempa {#1}\def\@tempb {#2}\def\@tempc
  {#3}\ifx \@tempc \@empty \let \@tempc \@tempb \let \@tempb \@tempa \fi \ifx
  \@tempb \@empty \def\@tempb {arXiv}\fi \@ifundefined
  {mn@eprint@\@tempb}{\@tempb:\@tempc}{\expandafter \expandafter \csname
  mn@eprint@\@tempb\endcsname \expandafter{\@tempc}}}

\bibitem[\protect\citeauthoryear{{Bower} et~al.,}{{Bower} et~al.}{2014}]{bow14}
{Bower} G.~C.,  et~al., 2014, \mn@doi [\apjl] {10.1088/2041-8205/780/1/L2},
  \href {http://adsabs.harvard.edu/abs/2014ApJ...780L...2B} {780, L2}

\bibitem[\protect\citeauthoryear{{Bower} et~al.,}{{Bower} et~al.}{2015}]{bow15}
{Bower} G.~C.,  et~al., 2015, \mn@doi [\apj] {10.1088/0004-637X/798/2/120},
  \href {http://adsabs.harvard.edu/abs/2015ApJ...798..120B} {798, 120}

\bibitem[\protect\citeauthoryear{{Camilo}, {Ransom}, {Halpern}, {Reynolds},
  {Helfand}, {Zimmerman}  \& {Sarkissian}}{{Camilo} et~al.}{2006}]{cam06}
{Camilo} F.,  {Ransom} S.~M.,  {Halpern} J.~P.,  {Reynolds} J.,  {Helfand}
  D.~J.,  {Zimmerman} N.,   {Sarkissian} J.,  2006, \mn@doi [\nat]
  {10.1038/nature04986}, \href
  {http://adsabs.harvard.edu/abs/2006Natur.442..892C} {442, 892}

\bibitem[\protect\citeauthoryear{{Camilo} et~al.,}{{Camilo}
  et~al.}{2007a}]{cam07a}
{Camilo} F.,  et~al., 2007a, \mn@doi [\apj] {10.1086/518226}, \href
  {http://adsabs.harvard.edu/abs/2007ApJ...663..497C} {663, 497}

\bibitem[\protect\citeauthoryear{{Camilo}, {Ransom}, {Halpern}  \&
  {Reynolds}}{{Camilo} et~al.}{2007b}]{cam07b}
{Camilo} F.,  {Ransom} S.~M.,  {Halpern} J.~P.,   {Reynolds} J.,  2007b,
  \mn@doi [\apjl] {10.1086/521826}, \href
  {http://adsabs.harvard.edu/abs/2007ApJ...666L..93C} {666, L93}

\bibitem[\protect\citeauthoryear{{Camilo} et~al.,}{{Camilo}
  et~al.}{2007c}]{cam07c}
{Camilo} F.,  et~al., 2007c, \mn@doi [\apj] {10.1086/521548}, \href
  {http://adsabs.harvard.edu/abs/2007ApJ...669..561C} {669, 561}

\bibitem[\protect\citeauthoryear{{Camilo}, {Reynolds}, {Johnston}, {Halpern}
  \& {Ransom}}{{Camilo} et~al.}{2008}]{cam08}
{Camilo} F.,  {Reynolds} J.,  {Johnston} S.,  {Halpern} J.~P.,   {Ransom}
  S.~M.,  2008, \mn@doi [\apj] {10.1086/587054}, \href
  {http://adsabs.harvard.edu/abs/2008ApJ...679..681C} {679, 681}

\bibitem[\protect\citeauthoryear{{Carter} et~al.,}{{Carter}
  et~al.}{2012}]{car12}
{Carter} M.,  et~al., 2012, \mn@doi [\aap] {10.1051/0004-6361/201118452}, \href
  {http://adsabs.harvard.edu/abs/2012A%26A...538A..89C} {538, A89}

\bibitem[\protect\citeauthoryear{{Cordes} \& {Lazio}}{{Cordes} \&
  {Lazio}}{1991}]{cl91}
{Cordes} J.~M.,  {Lazio} T.~J.,  1991, \mn@doi [\apj] {10.1086/170261}, \href
  {http://adsabs.harvard.edu/abs/1991ApJ...376..123C} {376, 123}

\bibitem[\protect\citeauthoryear{{Duncan} \& {Thompson}}{{Duncan} \&
  {Thompson}}{1992}]{dt92}
{Duncan} R.~C.,  {Thompson} C.,  1992, \mn@doi [\apjl] {10.1086/186413}, \href
  {http://adsabs.harvard.edu/abs/1992ApJ...392L...9D} {392, L9}

\bibitem[\protect\citeauthoryear{{Eatough} et~al.,}{{Eatough}
  et~al.}{2013}]{eat13}
{Eatough} R.~P.,  et~al., 2013, \mn@doi [\nat] {10.1038/nature12499}, \href
  {http://adsabs.harvard.edu/abs/2013Natur.501..391E} {501, 391}

\bibitem[\protect\citeauthoryear{{Hankins}, {Rankin}  \& {Eilek}}{{Hankins}
  et~al.}{2009}]{hank09}
{Hankins} T.~H.,  {Rankin} J.~M.,   {Eilek} J.~A.,  2009, in astro2010: The
  Astronomy and Astrophysics Decadal Survey.

\bibitem[\protect\citeauthoryear{{Kramer}}{{Kramer}}{1997}]{ckra97}
{Kramer} C.,  1997, Technical report, Calibration of spectral line data at the
  IRAM 30m radio telescope.
IRAM

\bibitem[\protect\citeauthoryear{{Kramer}, {Xilouris}, {Jessner}, {Wielebinski}
   \& {Timofeev}}{{Kramer} et~al.}{1996}]{kra96}
{Kramer} M.,  {Xilouris} K.~M.,  {Jessner} A.,  {Wielebinski} R.,   {Timofeev}
  M.,  1996, \aap, \href {http://adsabs.harvard.edu/abs/1996A%26A...306..867K}
  {306, 867}

\bibitem[\protect\citeauthoryear{{Kramer}, {Jessner}, {Doroshenko}  \&
  {Wielebinski}}{{Kramer} et~al.}{1997}]{kra97b}
{Kramer} M.,  {Jessner} A.,  {Doroshenko} O.,   {Wielebinski} R.,  1997, \apj,
  \href {http://adsabs.harvard.edu/abs/1997ApJ...488..364K} {488, 364}

\bibitem[\protect\citeauthoryear{{Kramer}, {Stappers}, {Jessner}, {Lyne}  \&
  {Jordan}}{{Kramer} et~al.}{2007}]{kra07}
{Kramer} M.,  {Stappers} B.~W.,  {Jessner} A.,  {Lyne} A.~G.,   {Jordan} C.~A.,
   2007, \mn@doi [\mnras] {10.1111/j.1365-2966.2007.11622.x}, \href
  {http://adsabs.harvard.edu/abs/2007MNRAS.377..107K} {377, 107}

\bibitem[\protect\citeauthoryear{{Kravchenko}, {Cotton}, {Yusef-Zadeh}  \&
  {Kovalev}}{{Kravchenko} et~al.}{2016}]{krav16}
{Kravchenko} E.~V.,  {Cotton} W.~D.,  {Yusef-Zadeh} F.,   {Kovalev} Y.~Y.,
  2016, \mn@doi [\mnras] {10.1093/mnras/stw304}, \href
  {http://adsabs.harvard.edu/abs/2016MNRAS.458.4456K} {458, 4456}

\bibitem[\protect\citeauthoryear{{Lazaridis}, {Jessner}, {Kramer}, {Stappers},
  {Lyne}, {Jordan}, {Serylak}  \& {Zensus}}{{Lazaridis} et~al.}{2008}]{laz08}
{Lazaridis} K.,  {Jessner} A.,  {Kramer} M.,  {Stappers} B.~W.,  {Lyne} A.~G.,
  {Jordan} C.~A.,  {Serylak} M.,   {Zensus} J.~A.,  2008, \mn@doi [\mnras]
  {10.1111/j.1365-2966.2008.13794.x}, \href
  {http://adsabs.harvard.edu/abs/2008MNRAS.390..839L} {390, 839}

\bibitem[\protect\citeauthoryear{{Levin} et~al.,}{{Levin} et~al.}{2010}]{lev10}
{Levin} L.,  et~al., 2010, \mn@doi [\apjl] {10.1088/2041-8205/721/1/L33}, \href
  {http://adsabs.harvard.edu/abs/2010ApJ...721L..33L} {721, L33}

\bibitem[\protect\citeauthoryear{{Levin} et~al.,}{{Levin} et~al.}{2012}]{lev12}
{Levin} L.,  et~al., 2012, \mn@doi [\mnras] {10.1111/j.1365-2966.2012.20807.x},
  \href {http://adsabs.harvard.edu/abs/2012MNRAS.422.2489L} {422, 2489}

\bibitem[\protect\citeauthoryear{{L{\"o}hmer}, {Jessner}, {Kramer},
  {Wielebinski}  \& {Maron}}{{L{\"o}hmer} et~al.}{2008}]{loeh08}
{L{\"o}hmer} O.,  {Jessner} A.,  {Kramer} M.,  {Wielebinski} R.,   {Maron} O.,
  2008, \mn@doi [\aap] {10.1051/0004-6361:20066806}, \href
  {http://adsabs.harvard.edu/abs/2008A%26A...480..623L} {480, 623}

\bibitem[\protect\citeauthoryear{{Lynch}, {Archibald}, {Kaspi}  \&
  {Scholz}}{{Lynch} et~al.}{2015}]{lynch15}
{Lynch} R.~S.,  {Archibald} R.~F.,  {Kaspi} V.~M.,   {Scholz} P.,  2015,
  \mn@doi [\apj] {10.1088/0004-637X/806/2/266}, \href
  {http://adsabs.harvard.edu/abs/2015ApJ...806..266L} {806, 266}

\bibitem[\protect\citeauthoryear{{Maron}, {Kijak}, {Kramer}  \&
  {Wielebinski}}{{Maron} et~al.}{2000}]{mar2000}
{Maron} O.,  {Kijak} J.,  {Kramer} M.,   {Wielebinski} R.,  2000, \mn@doi
  [\aaps] {10.1051/aas:2000298}, \href
  {http://adsabs.harvard.edu/abs/2000A%26AS..147..195M} {147, 195}

\bibitem[\protect\citeauthoryear{{Melrose} \& {Yuen}}{{Melrose} \&
  {Yuen}}{2016}]{melrose2016}
{Melrose} D.~B.,  {Yuen} R.,  2016, \mn@doi [Journal of Plasma Physics]
  {10.1017/S0022377816000398}, \href
  {http://adsabs.harvard.edu/abs/2016JPlPh..82b6302M} {82, 635820202}

\bibitem[\protect\citeauthoryear{{Michel}}{{Michel}}{1978}]{mich78}
{Michel} F.~C.,  1978, \mn@doi [\apj] {10.1086/155995}, \href
  {http://adsabs.harvard.edu/abs/1978ApJ...220.1101M} {220, 1101}

\bibitem[\protect\citeauthoryear{{Michel}}{{Michel}}{1982}]{mich82}
{Michel} F.~C.,  1982, \mn@doi [Reviews of Modern Physics]
  {10.1103/RevModPhys.54.1}, \href
  {http://adsabs.harvard.edu/abs/1982RvMP...54....1M} {54, 1}

\bibitem[\protect\citeauthoryear{{Morris}, {Graham}, {Sieber}, {Bartel}  \&
  {Thomasson}}{{Morris} et~al.}{1981}]{morr1981}
{Morris} D.,  {Graham} D.~A.,  {Sieber} W.,  {Bartel} N.,   {Thomasson} P.,
  1981, \aaps, \href {http://adsabs.harvard.edu/abs/1981A%26AS...46..421M} {46,
  421}

\bibitem[\protect\citeauthoryear{{Morris} et~al.,}{{Morris}
  et~al.}{1997}]{morr97}
{Morris} D.,  et~al., 1997, \aap, \href
  {http://adsabs.harvard.edu/abs/1997A%26A...322L..17M} {322, L17}

\bibitem[\protect\citeauthoryear{{Olausen} \& {Kaspi}}{{Olausen} \&
  {Kaspi}}{2014}]{ok14}
{Olausen} S.~A.,  {Kaspi} V.~M.,  2014, \mn@doi [\apjs]
  {10.1088/0067-0049/212/1/6}, \href
  {http://adsabs.harvard.edu/abs/2014ApJS..212....6O} {212, 6}

\bibitem[\protect\citeauthoryear{{Pe\~{n}alver}}{{Pe\~{n}alver}}{2012}]{pen12}
{Pe\~{n}alver} J.,  2012, Technical report, Antenna Technical Works.
IRAM

\bibitem[\protect\citeauthoryear{{Pennucci} et~al.,}{{Pennucci}
  et~al.}{2015}]{pen15}
{Pennucci} T.~T.,  et~al., 2015, \mn@doi [\apj] {10.1088/0004-637X/808/1/81},
  \href {http://adsabs.harvard.edu/abs/2015ApJ...808...81P} {808, 81}

\bibitem[\protect\citeauthoryear{{Shannon} \& {Johnston}}{{Shannon} \&
  {Johnston}}{2013}]{sj13}
{Shannon} R.~M.,  {Johnston} S.,  2013, \mn@doi [\mnras]
  {10.1093/mnrasl/slt088}, \href
  {http://adsabs.harvard.edu/abs/2013MNRAS.435L..29S} {435, L29}

\bibitem[\protect\citeauthoryear{{Spitler} et~al.,}{{Spitler}
  et~al.}{2014}]{spi14}
{Spitler} L.~G.,  et~al., 2014, \mn@doi [\apjl] {10.1088/2041-8205/780/1/L3},
  \href {http://adsabs.harvard.edu/abs/2014ApJ...780L...3S} {780, L3}

\bibitem[\protect\citeauthoryear{{Thompson} \& {Duncan}}{{Thompson} \&
  {Duncan}}{1995}]{td95}
{Thompson} C.,  {Duncan} R.~C.,  1995, \mnras, \href
  {http://adsabs.harvard.edu/abs/1995MNRAS.275..255T} {275, 255}

\bibitem[\protect\citeauthoryear{{Thompson} \& {Duncan}}{{Thompson} \&
  {Duncan}}{1996}]{td96}
{Thompson} C.,  {Duncan} R.~C.,  1996, \mn@doi [\apj] {10.1086/178147}, \href
  {http://adsabs.harvard.edu/abs/1996ApJ...473..322T} {473, 322}

\bibitem[\protect\citeauthoryear{{Torne} et~al.,}{{Torne} et~al.}{2015}]{tor15}
{Torne} P.,  et~al., 2015, \mn@doi [\mnras] {10.1093/mnrasl/slv063}, \href
  {http://adsabs.harvard.edu/abs/2015MNRAS.451L..50T} {451, L50}

\bibitem[\protect\citeauthoryear{{Wielebinski}, {Jessner}, {Kramer}  \&
  {Gil}}{{Wielebinski} et~al.}{1993}]{wiele93}
{Wielebinski} R.,  {Jessner} A.,  {Kramer} M.,   {Gil} J.~A.,  1993, \aap,
  \href {http://adsabs.harvard.edu/abs/1993A%26A...272L..13W} {272, L13}

\bibitem[\protect\citeauthoryear{{Xilouris}, {Kramer}, {Jessner}, {Wielebinski}
   \& {Timofeev}}{{Xilouris} et~al.}{1996}]{xil96}
{Xilouris} K.~M.,  {Kramer} M.,  {Jessner} A.,  {Wielebinski} R.,   {Timofeev}
  M.,  1996, \aap, \href {http://adsabs.harvard.edu/abs/1996A%26A...309..481X}
  {309, 481}

\makeatother
\end{thebibliography}







\bsp	
\label{lastpage}
\end{document}